\documentclass[12pt]{article}
\usepackage{graphics}
\usepackage{graphicx}

\usepackage{geometry}
\begin{document}

\rightline{Preprint RM3-TH/06-19}

\vspace{1cm}

\begin{center}

\huge{Leading and higher twists in proton, neutron and deuteron unpolarized
structure functions\footnote{To appear in the Proceedings of the $IV^{th}$ 
International Conference on {\em Quark and Nuclear Physics} (QNP06), Madrid 
(Spain), June 5-10, 2006.}}

\vspace{1cm}

\Large{Silvano Simula}

\vspace{0.5cm}

\normalsize{Istituto Nazionale di Fisica Nucleare, Sezione di Roma 3 \\ 
Via della Vasca Navale 84, I-00146 Roma, Italy}

\end{center}

\vspace{0.5cm}

\begin{abstract}
We summarize the results of a recent global analysis of proton and deuteron 
$F_2$ structure function world data performed over a large range of kinematics, 
including recent measurements done at JLab with the CLAS detector. 
From these data the lowest moments ($n \leq 10$) of the unpolarized structure 
functions are determined with good statistics and systematics. 
The $Q^2$ evolution of the extracted moments is analyzed in terms of an 
OPE based twist expansion, taking into account soft-gluon effects at large $x$.
A clean separation among the Leading and Higher-Twist terms is achieved. 
By combining proton and deuteron measurements the lowest moments of the 
neutron $F_2$ structure function are determined and its leading twist 
term is extracted.
Particular attention is paid to nuclear effects in the deuteron, which 
become increasingly important for the higher moments. Our results for 
the non-singlet, isovector ($p - n$) combination of the leading twist 
moments are used to test recent lattice simulations.
We also determine the lowest few moments of the higher twist contributions, 
and find these to be approximately isospin independent, suggesting the 
possible dominance of ud correlations over uu and dd in the nucleon.

\end{abstract}

\newpage

\pagestyle{plain}

In the past few years, thanks to the performance of the CLAS detector 
of the Hall B at Jefferson Lab, the unpolarized structure functions 
$F_2(x, Q^2)$ of both proton and deuteron have been measured precisely 
in a wide continuous interval of values of the Bjorken variable $x$ and 
of the squared four-momentum transfer $Q^2$ \cite{proton,deuteron}.
In particular, the $F_2$ structure functions have been extracted over 
the whole resonance region ($W \leq 2.5~GeV$) below $Q^2 \simeq 4.5 
~(GeV/c)^2$. These measurements, together with existing world data, 
have allowed the determination of the moments $M_n(Q^2)$ of the structure 
function $F_2$ up to $n = 10$, drastically reducing the uncertainties 
related to data interpolation and extrapolation as well as providing the 
most accurate evaluation of the $Q^2$-dependence of the moments starting 
from $Q^2 \simeq 0.1 ~(0.5)~(GeV/C)^2$ for proton (deuteron) up to $Q^2 
\simeq 100~(GeV/c)^2$.

A very powerful tool for analyzing in QCD the $Q^2$-dependence of the 
moments $M_n(Q^2)$ is represented by the Operator Product Expansion (OPE). 
Through OPE the moments are given as a series expansion in terms of matrix 
elements of local operators, namely
 \begin{equation}
    M_n(Q^2) = \sum_{\tau = 2}^{\infty} E_{n \tau}(\mu, Q^2) ~ O_{n \tau}(\mu) 
    \biggl(\frac{1}{Q^2}\biggr)^{{1 \over 2}(\tau - 2)},
    \label{eq:Mn}
 \end{equation}
where $\mu$ is the renormalization scale, $O_{n \tau}(\mu)$ is the reduced 
matrix element of local operators with definite spin $n$ and twist $\tau$, 
related to the non-perturbative structure of the target, and $E_{n \tau}(\mu, 
Q^2)$ is a short-distance coefficient calculable in pQCD thanks to the 
asymptotic freedom.

The first term of the series (\ref{eq:Mn}) corresponds to the Leading Twist 
(LT) and is determined by single-parton distributions in the target.
Subsequent terms are the so-called Higher Twists (HT's), which depend
on the interactions among partons (i.e., on multi-parton correlations). 
Their determination is considerably more challenging both experimentally 
and theoretically.

In Refs.~\cite{proton} and \cite{deuteron} both the LT and the HT's (up 
to twist-6) have been extracted from the proton and deuteron moments, 
respectively. Two important features of our twist analysis should be 
mentioned: 1) the analyzed moments are Nachtmann moments, and not 
Cornwall-Norton ones; 2) the perturbative evolution of the LT is 
calculated using soft-gluon resummation techniques. 

\begin{figure}[htb]
\includegraphics[bb=0.5cm 14cm 22cm 29cm, scale=0.75]{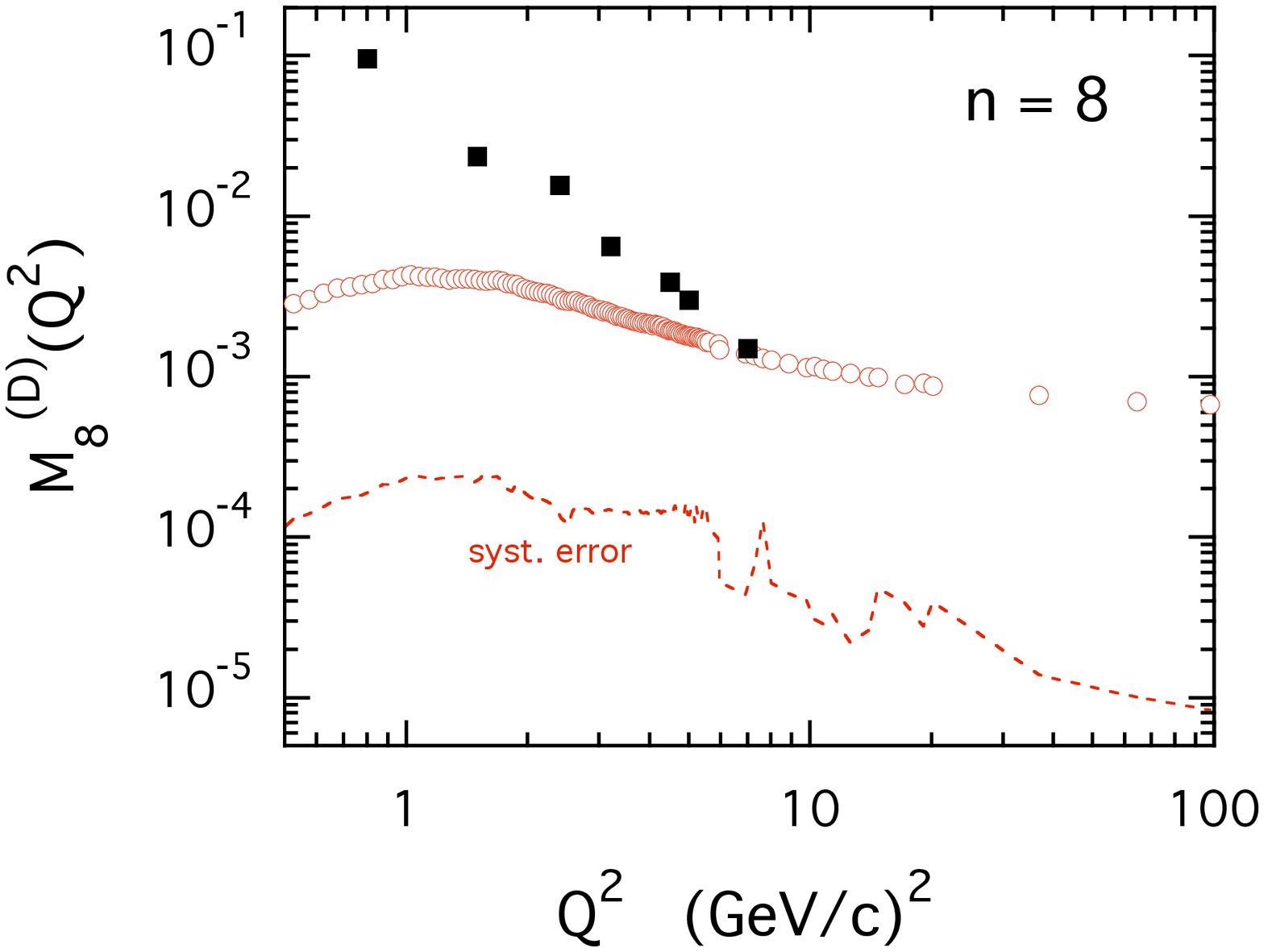}
\caption{\it Moment $M_8^{(D)}(Q^2)$ of the deuteron versus $Q^2$. Open dots are 
the results of Ref.~\cite{deuteron} using the Nachtmann definition of the 
moments, while full squares correspond to the CN moment as determined in 
Ref.~\cite{HallC}. The reported systematic errors have been precisely 
estimated only in Ref.~\cite{deuteron}.}
\label{fig:M8D}
\end{figure}

We have adopted the Nachtmann definition of the moments in order to get 
rid of the target-mass corrections, which are power corrections of kinematical 
origin and affect the Cornwall-Norton (CN) definition of the moments. 
The use of Nachtmann moments allows to extract in a clean way dynamical 
HT's, which are the only ones related to multi-parton correlations.
The target-mass corrections can be subtracted in principle from the CN 
moments, but this procedure requires a precise knowledge of the LT moments 
at large $n$. Moreover, the contribution of target-mass corrections is 
sizable just in the region of $Q^2 \simeq$ few $(GeV/c)^2$, where HT's are 
expected to show up.
This point is clearly illustrated in Fig.~\ref{fig:M8D}, where the Nachtmann 
moment $M_8^{(D)}(Q^2)$ is compared with the corresponding CN one as 
determined in Ref.~\cite{HallC}. It is also worthwhile to note the much 
larger number of very accurate data points obtained thanks to the CLAS 
detector.

As for the perturbative evolution of the LT, the presence of large logarithms 
in the coefficient function at large $n$ requires to go beyond any fixed-order 
approximation. As shown in Ref.~\cite{SGR} for a reliable extraction of both 
LT and HT's it is crucial to apply the resummation of soft gluons. Moreover, 
for internal consistency the value of $\alpha_s(M_Z^2)$ has been extracted 
in Ref.~\cite{alphaS} by analyzing the large $Q^2$ behavior of the proton 
moments of Ref.~\cite{proton} at the next-to-leading logarithmic accuracy 
used for the soft-gluon resummation.

The extracted LT components of proton and deuteron moments can be combined 
to form the LT moments of the neutron $F_2$ structure function.
In the Impulse Approximation the deuteron structure function can be 
written as the sum of two terms: $F_2^{(D)} = F_2^{(D, conv.)} + 
F_2^{(D, off)}$, where the first term is given by a convolution of the 
nucleon structure function and the nucleon momentum distribution function 
in the deuteron, $f^{(D)}$, while the second term represents nuclear off-shell 
plus relativistic corrections\footnote{For the moment with $n = 2$ shadowing 
and meson-exchange-current should be also considered (see Ref.~\cite{neutron}).}. 
Thus, the neutron moments can be obtained from:
 \begin{equation}
    \label{eq:nucl_cor}
     M_n^{(n)}(Q^2) = 2M_n^{(D)}(Q^2) \frac{1 - \Delta_n^{(off)}} {N_n^{(D)}} 
     - M_n^{(p)}(Q^2) ~,
 \end{equation}
where $N_n^{(D)}$ is the moment of the function $f^{(D)}$ and $\Delta_n^{(off)}$ 
represents the nuclear off-shell correction. 
Thus the quantity $N_n^{(D)} / [1 - \Delta_n^{(off)}]$ is the global nuclear 
correction factor, i.e. the EMC effect of the deuteron in moment space. 
The nuclear corrections and their uncertainties have been carefully estimated 
in Ref.~\cite{neutron}.

\begin{figure}[htb]
\includegraphics[bb=0.5cm 15cm 22cm 29cm, scale=0.75]{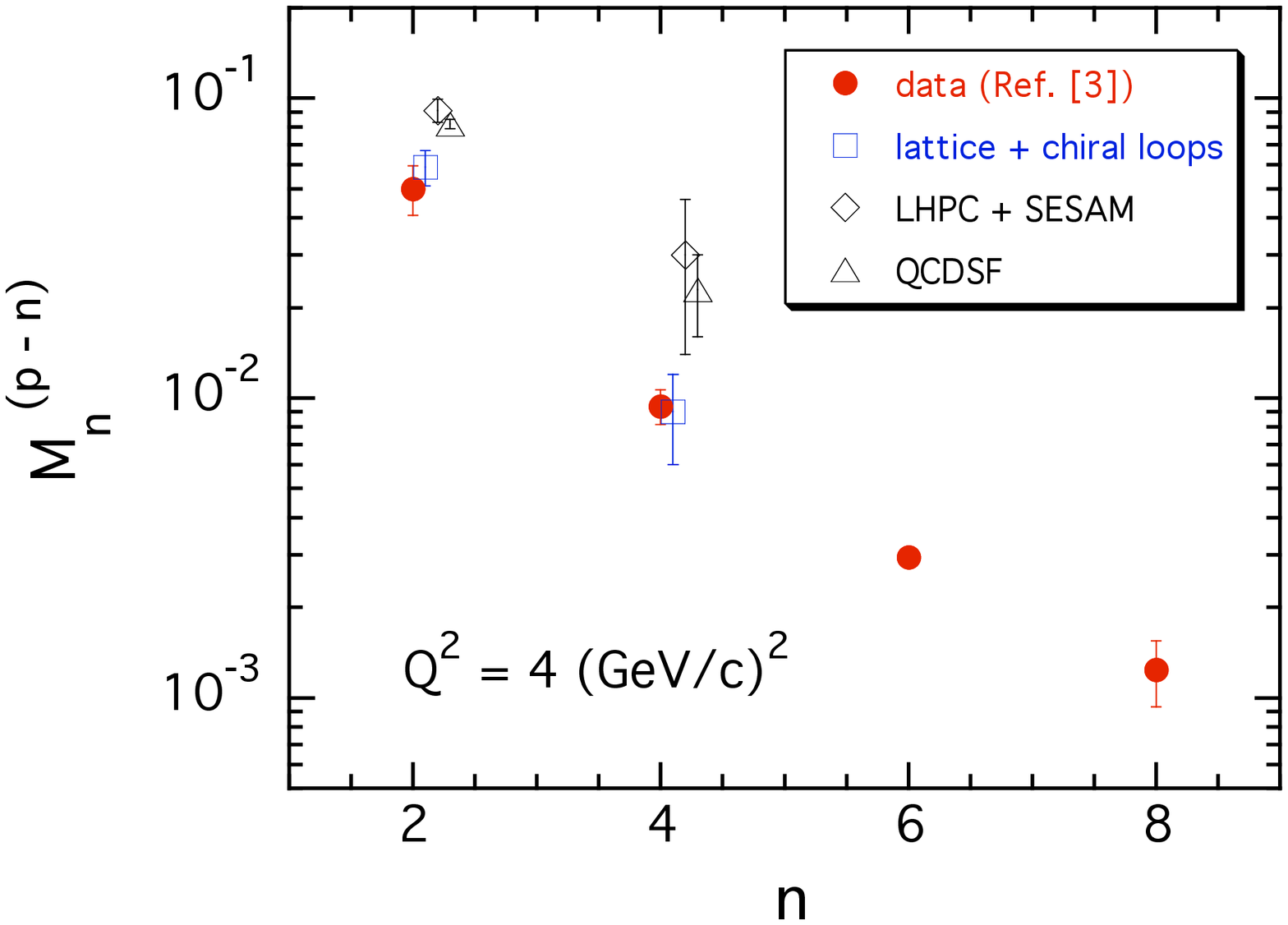}
\caption{\it Isovector ($p - n$) moments of the nucleon $F_2$ structure function
compared with lattice QCD simulations at $Q^2 = 4~(GeV/c)^2$. Full dots 
are the data from Ref.~\cite{neutron}, open diamonds and triangles 
correspond to the lattice results from Refs.~\cite{Dolgov} and 
\cite{Gockeler}, obtained assuming a linear extrapolation in the quark 
masses, and open squares are the results of Ref.~\cite{Detmold}, where 
chiral loop effects are considered.}
\label{fig:lattice}
\end{figure}

The extracted neutron LT moments can be combined with the proton LT ones to 
form the non-singlet, isovector ($p - n$) LT moments of the nucleon $F_2$ 
structure function. Our findings are compared against lattice QCD simulations 
in Fig.~\ref{fig:lattice}. While a linear extrapolation of the lattice results 
to the physical pion mass overestimates our data significantly, the results 
obtained using the extrapolation method of Ref.~\cite{Detmold} are in much 
better agreement. Our results for higher moments exhibit a remarkable 
precision, and it would be valuable to compare them with corresponding 
lattice moments, because the effects of chiral loops are expected to be 
suppressed at large $n$. 

\begin{figure}[htb]
\includegraphics[bb=0cm 5cm 22cm 24cm, scale=0.7]{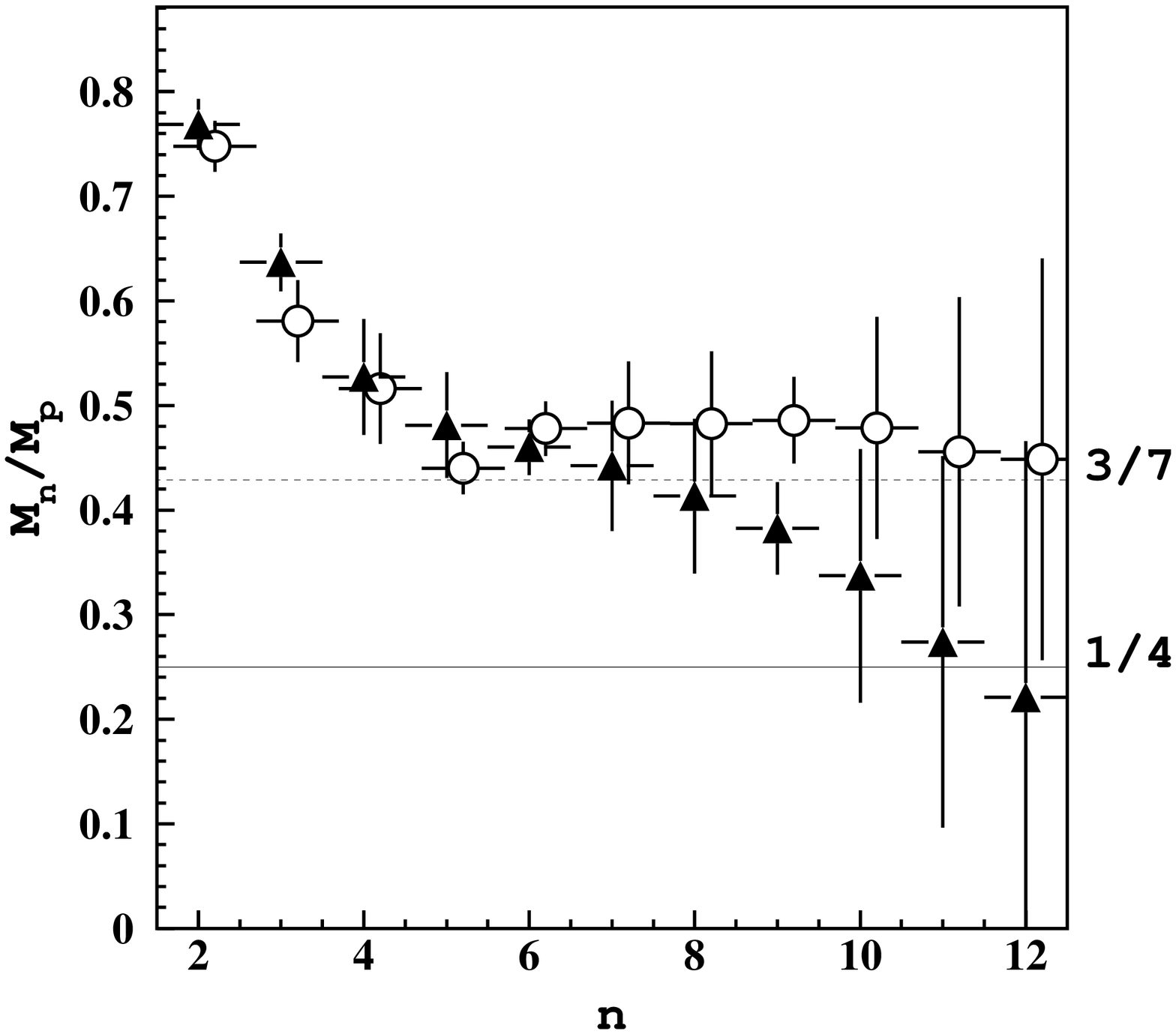}
\caption{\it Ratio of neutron to proton moments versus $n$: full triangles and 
open dots corresponds to the results of Ref.~\cite{neutron} obtained with 
and without the nuclear corrections, respectively. The solid (dashed) line 
indicates the scenario where $d / u \to 0~(1/5)$ for $x \to 1$.}
\label{fig:nucl_cor}
\end{figure}

Another interesting issue is the behavior of the neutron to proton structure 
function ratio $R(x) \equiv F_2^n(x) / F_2^p(x) $ in the limit $x \to 1$. 
This ratio can be easily related to the corresponding $d / u$ ratio of parton 
distributions. The ``standard'' value $R(x = 1) = 1/4$ is used in most of the 
parton distribution fits and corresponds to a vanishing $d / u$ ratio, while 
the value $R(x = 1) = 3/7$ is expected from pQCD arguments and corresponds to 
a limiting value of 1/5 for the $d / u$ ratio. As shown in Ref.~\cite{neutron} 
the ratio of neutron to proton moments, $M_n^n/M_n^p$, is largely independent 
of $Q^2$ and is very sensitive at large $n$ to the large-$x$ behavior of $R(x)$. 
Our results are shown in Fig.~\ref{fig:nucl_cor}. The trend is clearly towards 
the ``standard'' value 1/4 as $n$ increases, although the precision of the 
data does not exclude higher limiting values.
From Fig.~\ref{fig:nucl_cor} one can also see the impact of the nuclear 
corrections on the deuteron moments at large $n$: for $n = 12$, which
corresponds to an average value of $x$ around 0.75, the nuclear correction
corresponds to a reduction factor of $\simeq 2$, though within large 
uncertainties.

\begin{figure}[htb]
\includegraphics[bb=0cm 5cm 22cm 24cm, scale=0.7]{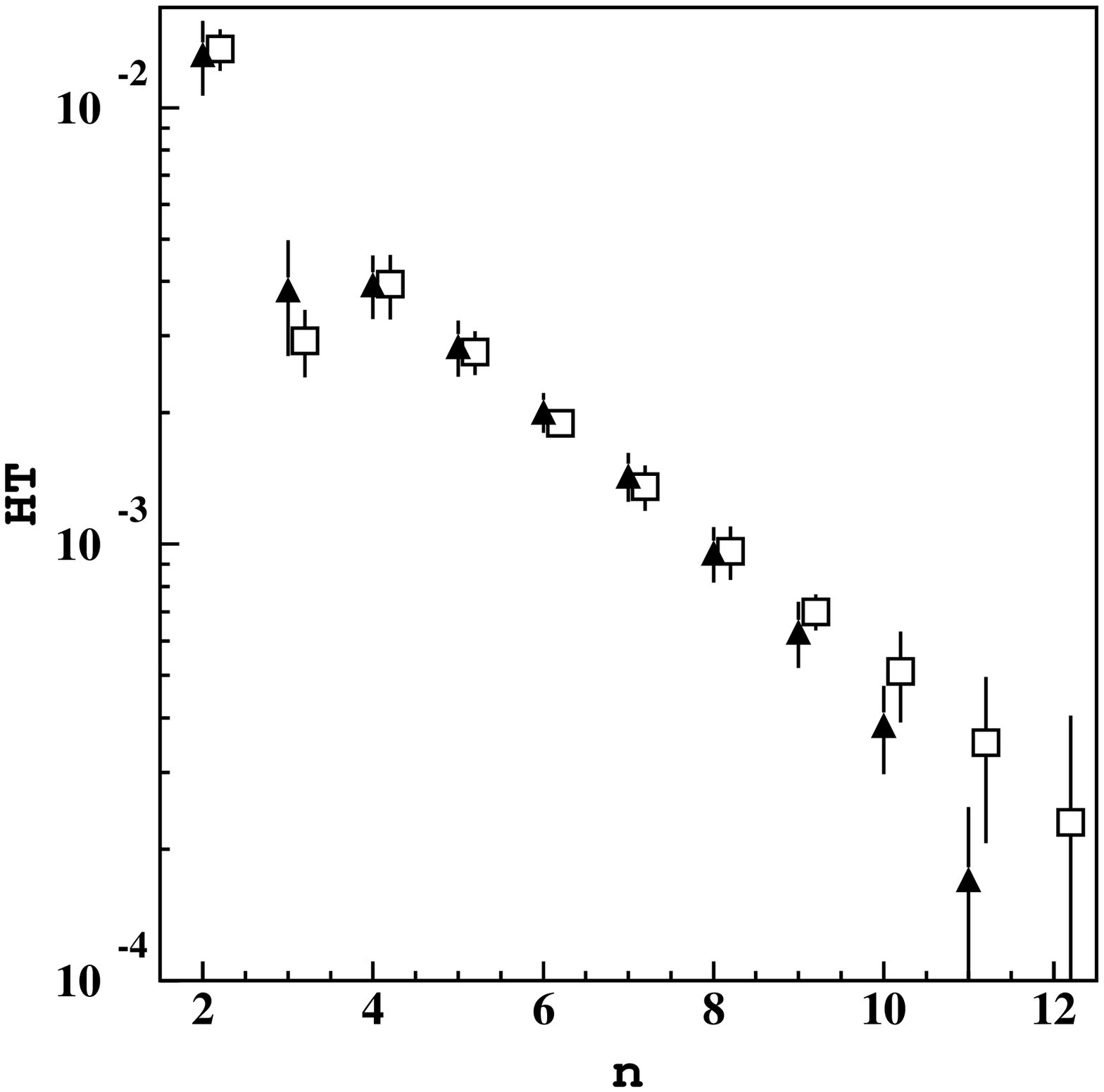}
\caption{\it Total higher twist contribution to the proton (filled triangles) and 
deuteron (open squares) moments evaluated at $Q^2 = 2~(GeV/c)^2$.}
\label{fig:hts}
\end{figure}

Once the LT contribution to moments is determined, one can study the isospin 
dependence of the HT contribution. Assuming that final state interactions and 
meson exchange currents have negligible impact on the HT's above $Q^2 \simeq 
1~(GeV/c)^2$, the same nuclear corrections used for the LT can be applied to 
the HT's. In this way the total HT contributions to proton moments and to the 
corrected deuteron moments can be compared, as done in Fig.~\ref{fig:hts}.
The comparison indicates clearly that the total HT contribution is almost 
independent of isospin, suggesting the possible dominance of ud correlations 
over uu and dd in the nucleon \cite{neutron}. Moreover, the isovector 
combination ($p - n$) of the $F_2$ structure functions should be almost 
free from HT contributions within the present uncertainties.

In conclusion we have analyzed experimental data on proton and deuteron $F_2$ 
structure functions in order to determine their moments. Using the OPE we have 
extracted both leading and higher twist contributions.
By combining proton and deuteron moments we have determined the LT moments 
of the neutron $F_2$ structure function, paying particular attention to the 
issue of nuclear effects in the deuteron, which are increasingly important 
at large $n$.

The main results of our analysis can be summarized as follows:

\begin{itemize}

\item
the ratio of neutron to proton moments is consistent with $F_2^n / F_2^p 
\to 1/4$ as $x \to 1$, although the higher value of 3/7, suggested by pQCD 
arguments, cannot be excluded;

\item
the non-singlet moments are in good agreement with the lattice data 
\cite{Dolgov,Gockeler} only when the latter are extrapolated to physical 
quark masses after having taken into account chiral loops \cite{Detmold};

\item
the total contribution of HT's is found to be isospin independent, suggesting 
the possible dominance of ud correlations over uu and dd in the nucleon. 
This implies that in the isovector combination ($p - n$) of $F_2$ structure 
functions the HT's are almost consistent with zero within the present 
uncertainties.

\end{itemize}


\end{document}